\documentclass{ws-ijmpcs}

\newcommand{\rt}{{\mathbf{r}_T}}
\newcommand{\xt}{{\mathbf{x}_T}}
\newcommand{\ut}{{\mathbf{u}_T}}

\newcommand{\vt}{{\mathbf{v}_T}}
\newcommand{\bt}{{\mathbf{b}_T}}
\newcommand{\yt}{{\mathbf{y}_T}}
\newcommand{\zt}{{\mathbf{z}_T}}
\newcommand{\ud}{\, \mathrm{d}}
\newcommand{\nc}{{N_\mathrm{c}}}
\newcommand{\tr}{\, \mathrm{Tr} \, }
\newcommand{\as}{\alpha_{\mathrm{s}}}

\begin{document}


%
\catchline{}{}{}{}{}
%

\title{COMPUTING EARLY-TIME DYNAMICS IN HEAVY ION COLLISIONS: STATUS, PROBLEMS AND PROSPECTS
}

\author{RAJU VENUGOPALAN}

\address{Physics Department, Brookhaven National Laboratory\\
Upton, NY 11901, USA
\\
raju@bnl.gov}

\maketitle


\begin{abstract}
We discuss some recent developments towards a quantitative understanding of the production and early-time evolution of bulk quark-gluon matter in ultrarelativistic heavy ion collisions.

\end{abstract}

\ccode{PACS numbers: 11.25.Hf, 123.1K}

%

\section{Introduction}	

This talk summarizes some recent theoretical developments in high energy QCD, in particular those that deal with very high energy heavy ion collisions. Compared to the early days of studies of heavy ion collisions, the developments have been significant though outstanding problems remain.  One such development is a QCD effective field theory (EFT)  of high energy wavefunctions, the Color Glass Condensate (CGC)~[\refcite{CGC-reviews}], which provides a unified systematic non-perturbative (albeit weak coupling) approach to compute key features of e+A, p+A and A+A collisions. This framework has been applied successfully to describe inclusive and diffractive data at HERA, inclusive data from fixed target e+A experiments, singly and doubly inclusive deuteron-gold data from RHIC, the energy dependence and multiplicity distributions of p+p collisions up to the highest LHC energies, the centrality and multiplicity distributions in A+A collisions, and their energy dependence from RHIC to LHC. Long range rapidity correlations in p+p and A+A collisions arise naturally in the CGC; a noteworthy recent example is the ``ridge" in p+p collisions which was predicted in this formalism. While these comparisons to data are not fully quantitative, the economy and consistency of parameters, and the unified description of multi-particle production in deeply inelastic scattering and hadronic collisions, make the CGC an attractive candidate for a quantitative description of the formation and evolution of quark-gluon matter in heavy ion collisions. 

We shall first briefly discuss the many-body structure of high energy wavefunctions in the CGC. We next describe how these wavefunctions collide and decohere to produce quark-gluon matter called the Glasma. The Glasma, initially described by high occupation number fields, evolves in a highly non-trivial manner to likely form a thermalized Quark-Gluon Plasma (QGP). In the final section, we will outline a scenario whereby this process occurs, and shall discuss possible implications of these results.

\section{Before the collision: hadron/nuclear structure at high energies}

QCD in high energy Regge asymptotics is a dense many-body system of ``wee" gluons and sea quarks.  In the infinite momentum frame, gluons with transverse momenta $k_\perp \leq Q_S$ saturate phase space maximally, where $Q_S (x)$ is a dynamical saturation scale~[\refcite{GLR}] that grows with decreasing fractions $x$ of the longitudinal momentum of the hadron carried by the gluons. The properties of saturated gluons are described by the CGC EFT: the degrees of freedom are static color sources at large $x$, coupled to dynamical wee gluon fields at small $x$. Because of the large occupancy of wee gluons, the effective ground state wavefunction is a classical non-Abelian Weizs\"{a}cker-Williams field~[\refcite{MV}], whose properties at small $x$, on account of $Q_S\gg \Lambda_{\rm QCD}$, can be described in weak coupling. Renormalization group (RG) equations, derived from requiring that observables be independent of the separation in $x$ between sources and fields, lead to an infinite hierarchy of evolution equations in $x$, for n-point Wilson line correlators averaged over dense color fields in the nucleus.  For a physical observable defined by an average over all static color source configurations,
\begin{equation}
\langle {\cal O}\rangle_{_Y} \equiv \int [D\rho]\; W_{_Y}[\rho]\;
 {\cal O}[\rho] \; ,
\label{eq:CGC-avg}
\end{equation}
the gauge invariant weight functional of color sources $W_{_Y}[\rho]$ at rapidity $Y=\ln(x/x_0)\equiv \ln(x_0^-/x^-)$, where $x_0$ is the initial scale for small x evolution, satisfies the JIMWLK equation~[\refcite{JIMWLK}] $\partial W_{_Y}[\rho] / \partial Y = {\cal H}\,W_{_Y}[\rho]$. The energy evolution of the observable can then be expressed as 
\begin{equation}
{\partial \langle {\cal O}\rangle_{_Y}\over \partial Y} = \langle {\cal H}\;
 {\cal O}\rangle_{_Y} \; .
\label{eq:CGC-hier}
\end{equation}
The structure of  the JIMWLK Hamiltonian ${\cal H}$ is such that $\langle {\cal H}{\cal O}\rangle_{_Y}$ is an object distinct from $\langle {\cal O}\rangle_{_Y}$; one obtains an infinite hierarchy of evolution equations for operator expectation values $\langle{\cal O}\rangle_{_Y}$. 
Given appropriate initial conditions at large $x$, solutions of this Balitsky-JIMWLK hierarchy~[\refcite{Balitsky:1995ub,JIMWLK}] allow one to compute a wide range of multi-particle final states in deeply inelastic scattering (DIS) and hadronic collisions. 

The simplest example is that of the inclusive DIS structure functions 
$F_{2}$ and $F_{L}$, which are proportional to the forward 
scattering amplitude of a $q\bar{q}$  ``dipole'' on a nucleus. 
The forward dipole amplitude (dipole cross section) is given by 
\begin{eqnarray}
\sigma_\mathrm{dip.}(x, \rt) = 2\int \ud^2 \bt 
\times
\bigg< 1 
- \frac{1}{\nc}
\tr  V\left(\bt + \frac{\rt}{2}\right)
V^\dagger\left(\bt - \frac{\rt}{2}\right) \bigg> \, ,
\label{eq:dipole}
\end{eqnarray}
where $\rt = \xt -\yt$ is the transverse size of the dipole, 
$\bt = (\xt+\yt)/2$ is the impact parameter relative to the hadron and $V(\xt) = P\exp(ig \int dz^- \,{\rho(\xt,z^-)\over\nabla^2})$.  From eq.~(\ref{eq:CGC-hier}), one can show that the expectation value $D \equiv \langle \hat{D}\rangle$, with 
$\hat{D}(\xt-\yt) \equiv \frac{1}{\nc} \tr V(\xt) V^\dagger(\yt)$, obeys the Balitsky-JIMWLK equation that relates its energy dependence to the expectation value of a four-point operator, 
\begin{eqnarray} \label{eq:hierarchy}
 {\ud\over \ud Y} D (\xt - \yt) =  
{\nc\, \as \over 2\pi^2} 
\int_\zt  {\cal K}_{{\bf x}{\bf y} {\bf z}} 
\left\langle \hat{D} (\xt - \zt)\, \hat{D} (\zt - \yt) - \hat{D} 
(\xt - \yt)\right\rangle ,
\end{eqnarray}
where ${\cal K}_{{\bf x} {\bf y} {\bf z}}=(\xt - \yt)^2 / (\xt - \zt)^2 (\zt - \yt)^2$. 
In the large $\nc$ approximation, the expectation value of $\hat{D}^2$
factorizes--the resulting closed form expression  
\begin{eqnarray}
 {\ud\over \ud Y} D (\xt - \yt) =  
{\nc\, \as \over 2\pi^2} 
\int_\zt {\cal K}_{{\bf x} {\bf y} {\bf z}} 
\left[D (\xt - \zt)\, D (\zt - \yt) - D (\xt - \yt)\right].
\label{eq:2pt}
\end{eqnarray}
is known as the Balitsky-Kovchegov (BK) equation~[\refcite{Balitsky:1995ub,Kovchegov:1999yj}]. In addition to eq.~(\ref{eq:dipole}), the dipole correlator appears in a number of final states in both DIS and hadronic scattering; the BK equation for its energy evolution is successfully employed in phenomenological applications. 

For less inclusive observables, one encounters the expectation value $Q=\langle {\hat Q}\rangle$ of  the ``quadrupole" operator
\begin{equation}
{\hat Q}(\yt,\xt,\ut,\vt) = {1\over \nc} \tr \left(V^\dagger(\yt) V(\xt) V^\dagger (\ut) V (\vt)\right)  \, .
\label{eq:quad}
\end{equation}
Unlike $\langle {\hat D} {\hat D}\rangle$ in eq.~(\ref{eq:hierarchy}), it is not reducible to the product of dipoles even in the large $\nc$ and large $A$ approximations and is a novel universal correlator in high energy QCD~[\refcite{JalilianMarian:2004da,Dominguez:2011wm}], interesting
both from theoretical and phenomenological perspectives. Examples of where this quantity appears include small-$x$ di-jet production in e+A DIS~\cite{Dominguez:2011wm}, quark-antiquark pair production in hadronic collisions~[\refcite{Blaizot:2004wv}] and near-side long-range rapidity correlations~[\refcite{Dumitru:2010mv}]. Another interesting quantity~[\refcite{Dominguez:2011wm}] is a six-point correlator $S_6 \propto \langle {\hat Q} {\hat D}\rangle$ that is probed in forward di-hadron production $d+A\longrightarrow h_1\;h_2\;X$ in deuteron-gold collisions at RHIC.  The dominant underlying QCD process is the scattering of a large $x_1$ valence quark from the deuteron off small $x_2$ partons in the nuclear target, with the emission of a gluon from the valence quark either before or after the collision. The RHIC experiments show that the away-side peak in the di-hadron correlations is significantly broadened for central collisions at forward rapidities~[\refcite{Adare:2011sc,Braidot:2011zj}] as predicted in the CGC framework~[\refcite{Marquet:2007vb}] and confirmed in more detailed analyses~[\refcite{Albacete:2010pg,Tuchin:2009ve}]. These analyses however relied on 
simplified factorization assumptions that are not justified. The JIMWLK RG equations for  $Q$~[\refcite{Dominguez:2011gc}] and $S_6$~[\refcite{Dumitru:2010ak}] have been derived and computed numerically for particular configurations of these operators~[\refcite{Dumitru:2011vk}]. A remarkable result of these numerical simulations is that the initial condition and the RG evolution of these higher point correlators are well reproduced by expressions that are functions only of the dipole expectation value $D$--see ref.~[\refcite{Iancu:2011ns}] for a recent analytical interpretation of these results.

As this discussion suggests, significant progress has been made in understanding the RG evolution of novel multi-point correlations that comprise the structure of hadron wavefunctions at high energies and observables sensitive to these correlators in p+A and e+A collisions have been identified. An outstanding problem that is unresolved is the impact parameter dependence of high energy evolution.  Exclusive measurements at an e+A collider provide the best opportunity to shed empirical light on this topic~[\refcite{Boer:2011fh}].

\section{Creating bulk matter: The Glasma}

Before the collision, the effective ground states of the nuclei are coherent states described as classical fields with occupation numbers $O(1/g^2)$. 
These fields become time dependent immediately after the collision. Inclusive quantities, computed in the presence of these time dependent color fields, can be  expressed in terms of retarded propagators, thereby allowing real time computations with initial data at negative infinity~[\refcite{Gelis:2006yv,Gelis:2006cr}]. Other examples of similar non-perturbative dynamics include Schwinger's mechanism for electron-positron pair production in strong QED fields and Hawking radiation in the vicinity of the event horizon of a black hole. 

The dynamics of the Glasma~[\refcite{Lappi:2006fp}] created in heavy ion collsions is analogous to important aspects of early universe cosmology, one being an``Ekpyrotic" picture of the production of multiverses through the collision of branes; another powerful analogy is the dynamics of pre-heating that leads to thermalization and hydrodynamics of classical fields~[\refcite{Dusling:2011gk}]. In QCD, a classical description of Ekpyrosis is obtained by solving  Yang-Mills equations with static light front color  sources~[\refcite{Kovner:1995ts,Krasnitz:1999wc,Lappi:2003bi}]. At this classical level, energy dependence is introduced ``by hand" via the saturation scale $Q_S$. The energy dependence arises at next-to-leading (NLO) order from quantum fluctuations (of relative strength $g$) about the classical background of the nuclei. These naively sub-leading contributions are enhanced by logarithms, which give $\alpha_S\ln(x_{1,2}) \sim 1$ for small $x_{1,2}$ and have to be resummed to all orders in perturbation theory. In the strong field regime, there is an additional resummation $(g\rho_{1,2})^n$ at each order in the $\alpha_S$ expansion. These respectively radiative and multiple scattering contributions are generated by the JIMWLK RG equation for the weight functionals $W_{Y_{1(2)}}[\rho_{1(2)}]$ for each of the incoming nuclei. First computations of the inclusive multiplicities in heavy ion collisions by solving Yang-Mills equations with JIMWLK generated initial conditions have been performed recently~[\refcite{Lappi:2011ju}]. 

An important feature of the Glasma is that it generates long-range correlations that are localized on scales $\sim 1/Q_S$ in the transverse plane~[\refcite{Dumitru-GMV}]. The long range correlations evolve with energy and the the rapidity separation between correlated gluons~[\refcite{Dusling-GLV}]. This formalism predicted~[\refcite{Dumitru}] a near-side ``ridge" in high multiplicity collisions, an effect that was observed by the CMS collaboration~[\refcite{CMS-ridge}]. The systematics of the data are in qualitative agreement with the Glasma predictions~[\refcite{Dumitru-etal}]. In A+A collisions, the observed ridge  is a consequence of the boosting of long-range correlations in the final state by radial flow~[\refcite{Voloshin-Shuryak}]. The radial flow provides the near side angular collimation. The combination of flux tube structures in the initial state and radial flow give a good description of  RHIC and LHC data~[\refcite{Gavin-Moschelli,Werner-Kodama}]. While it would be interesting to discuss at length these long range correlations,  the rest of this talk will focus on the space-time evolution of the Glasma. 

Albeit causality requires that nuclei don't communicate before the collision, it is by no means assured in a weak coupling treatment; we are able to show formally that factorization of the contributions of the weight functionals $W[\rho_{1,2}]$ to inclusive quantities is obtained at leading logs in $x$ accuracy~[\refcite{Gelis:2008rw,Gelis:2008ad,Gelis:2008sz}]. These quantum modes are boost-invariant $p^\eta=0$ modes, where $p^\eta$ is Fourier conjugate to the space-time rapidity. In the process of Ekpyrosis, nuclear coherence is lost and $p^\eta\neq 0$ modes are generated. These modes are generically unstable~[\refcite{Romatschke:2005pm,Romatschke:2006nk,Romatschke:2005ag,Fukushima:2011nq}], and grow in an expanding system as $(\alpha_S\exp(2\sqrt{\mu \tau}))^n$, with $\mu \sim Q_S$,  where $n$ denotes the order in perturbation theory beyond the classical leading order contribution.  These ``leading instabilities" are comparable to the background field at $\tau\sim 1/Q_S$ and have to be resummed to all orders, leading to qualitatively different behavior. 

After factorization and resummation of leading logs in $x_{1,2}$ and leading instabilities $\alpha_S\exp(2\sqrt{\mu \tau})$, the energy-momentum tensor is expressed as 
\begin{equation}
\langle T^{\mu\nu}\rangle_{{\rm LLx + LInst.}} =\int [D\rho_1 D\rho_2]\;
W_{x_1}[\rho_1]\, W_{x_2}[\rho_2]  \int \!\! \big[{\cal D}\alpha\big]\,
F_0\big[\alpha\big]\; T_{_{\rm LO}}^{\mu\nu} [{\cal A}[\rho_1,\rho_2] + \alpha] (x)\; .
\label{eq:final-formula}
\end{equation}
The argument ${\cal A}\equiv ( A, E)$ denotes collectively the components of the classical fields and their canonically
conjugate momenta on the initial proper time surface; analytical expressions for these are available at
$\tau=0^+$~[\refcite{Kovner:1995ts,Krasnitz:1998ns}].  The initial spectrum of fluctuations $F_0\big[\alpha\big]$, Gaussian in the quantum fluctuations $\alpha$,  has a variance given by the small fluctuation propagator in the Glasma background field as $\tau\rightarrow 0^+$. In practice, the path integral in $\alpha$ is determined by solving the classical Yang-Mills equations repeatedly with the initial conditions at $\tau=0^+$ given by
\begin{equation}
{\bf A}_{\rm init.} = {\cal A}_{\rm init.} + \int d\mu_{_K}\;\Big[c_{_K}\,a_{_K}^\mu(x)+c_{_K}^*\,a_{_K}^{\mu*}(x)\Big] \, .
\label{eq:quantum}
\end{equation}
Here ${\bf A}$ collectively denotes the quantum fields and their canonical conjugate momenta. The  coefficients $c_{_K}$, with $K$ collectively denoting the quantum numbers labeling the basis of solutions, are random Gaussian-distributed complex numbers given by
\begin{eqnarray}
\left<c_{_K}c_{_{K^\prime}}^*\right>=\frac{N_{_K}}{2} \delta_{_{KK^\prime}}
\;\; ,\;\;\left<c_{_K}c_{_{K^\prime}}\right>=\left<c_{_K}^*c_{_{K^\prime}}^*\right>=0 \; .
\label{eq:random2}
\end{eqnarray}
Explicit expressions for the small fluctuations and their conjugate momenta, denoted here by $a_K^\mu (x)$ were obtained in ref.~[\refcite{Dusling:2011rz}]. The inner product of these solutions satisfies the orthogonality condition $(a_{_K}\big|a_{_{K^\prime}}\big)=N_{_K}\,\delta_{_{KK^\prime}}$ with the measure $d\mu_{_K}$ (a mix of integrals and discrete sums) that ensures $\int d\mu_{_K}\;N_{_K}\,\delta_{_{KK^\prime}}=1$. 

In ref.~[\refcite{Dusling:2011rz}], a numerical algorithm was outlined to compute eq.~(\ref{eq:final-formula}), thereby describing both Ekpyrosis and inflationary dynamics including essential leading quantum fluctuations. (We emphasize the formalism holds for any inclusive quantity in heavy-ion collisions including  parton energy loss and sphaleron transitions at early times.) We can now study how lumpy initial conditions for color charges $\rho_{1,2}^a$ in nuclear wavefunctions at the energy/rapidity of interest transform into the flow of matter.  The early universe analogy is becoming more robust experimentally with a ``WMAP-like" spectrum of spatial anisotropies resolved in heavy ion data~[\refcite{Sorensen:2011hm}].

We carried out an extensive study of the formalism outlined here for a scalar $\phi^4$ model which, among several  QCD-like features, has a spectrum of unstable quantum modes which are amplified by resonant interactions with the background field~[\refcite{Dusling:2010rm}]. In $\phi^4$ (and other scale invariant theories), the amplitude of the field is inversely proportional to its period of oscillation. Slightly different amplitudes, corresponding to different quantum seeds, lead to differing oscillation periods; a stochastic average over these leads to decoherence in the evolution.  A striking consequence is hydrodynamic flow with an ideal equation of state. Eq.~(\ref{eq:quantum})  is a  realization of Berry's conjecture~[\refcite{Berry}] which is believed to be necessary for thermalization~[\refcite{Srednicki}] of a quantum system. Thermalization and onset of quasi-particle dynamics have been studied in the scalar theory for a fixed box~[\refcite{Epelbaum:2011pc}]; similar studies are feasible in the expanding case and eventually in the QCD framework of ref.~[\refcite{Dusling:2011rz}]. A promising development~[\refcite{Hatta:2011ky}] is an attempt to connect the CGC power  counting approach here with a well developed 2PI formalism~[\refcite{Berges:2008mr,Reinosa:2009tc}] which has to potential to allow one to follow the evolution of the system in heavy ion collisions until thermalization. In the following section, we will describe a complementary approach based on the Boltzmann equation that describes how the highly occupied initial state approaches thermalization. 

\section{Space-time evolution of strongly interacting matter: from Glasma to Quark-Gluon Plasma}

The classical Yang-Mills energy-momentum tensor in the Glasma is of the form $T^{\mu\nu}={\rm diag}\,(\epsilon,\epsilon,\epsilon,-\epsilon)$, and therefore has a negative longitudinal pressure at very early times~[\refcite{Krasnitz:2002mn}]. The unstable quantum fluctuations we discussed above can however rapidly isotropize the energy-momentum tensor on very short time scales of order $1/Q_S$. An outstanding issue then is whether the system can resist the tendency of to fall out of isotropy again due to its rapid expansion into the vacuum. Another outstanding issue is whether the system can equilibrate to generate a Bose-Einstein distribution on the short time scales available. The ``bottom-up" scenario~[\refcite{Baier:2000sb}] outlined a systematic way whereby  hard elastic and inelastic collisions of produced gluons would drive the system towards equilibration.  In ref.~[\refcite{Arnold:2003rq}], it was demonstrated that anisotropy driven instabilities can significantly alter the bottom-up picture. Here, we will discuss an alternative scenario~[\refcite{Blaizot:2011xf}] where early time instabilities, as discussed previously, have already played a role, generating  a gluon density in the Glasma that is parametrically large compared to the value it should have in a system in thermal equilibrium with the same energy density. In such systems, the excess of gluons can be diluted by dynamical generation of a Bose-Einstein condensate, corresponding to a large occupation of the zero momentum mode, and/or by inelastic processes that in the long run tend to tame the particle excess. Until the latter begin to dominate, a transient Bose-Einstein condensate can exist, as in superfluids, with interesting consequences. 

We will assume that after times $\sim 1/Q_{\rm s}$, the Glasma can be described by  color singlet distributions for both the particle content and the condensate. We will study in kinetic approach the role of collisions in driving the overpopulated system to equilibration. We will assume that, at all times $t>1/Q_S$, the distribution function takes the form 
\begin{equation} 
f(p)\sim\frac{1}{\alpha_{\rm s}} \;\; {\rm for} \; p<\Lambda_{\rm s}, \qquad f(p)\sim
\frac{1}{\alpha_{\rm s}} \frac{\Lambda_{\rm s}}{\omega_{\bf p}} \;\; {\rm for} \;
\Lambda_{\rm s}<p<\Lambda, \qquad f(p)\sim 0\;\; {\rm for} \; \Lambda<p.
\end{equation} 
At $t\sim 1/Q_{\rm s}$, both scales $\Lambda_{\rm s}$ and $\Lambda$ coincide with $Q_{\rm s}$. As time progresses, the two scales separate, with $\Lambda_{\rm s}$ decreasing quickly, and $\Lambda$ evolving much more slowly. Thermalization is reached when $\Lambda_{\rm s}/\Lambda\sim \alpha_{\rm s}$, at which point, $f(\Lambda)$ becomes of order unity.

A more precise definition of these scales is obtained from the collision integral of the Boltzmann equation in the small angle
approximation, assuming $2\to 2$ elastic scattering and isotropy,
\begin{equation} 
\left. \frac{\partial
    f}{\partial t}\right|_{\rm coll}\sim
\frac{\Lambda_{\rm s}^2\Lambda}{p^2} \partial_p \left\{ p^2\left[
    \frac{df}{dp}+\frac{\alpha_{\rm s}}{\Lambda_{\rm s}} f(p)(1+f(p)) \right]
\right\}.  \end{equation} 
The fixed point solution of this equation is a
Bose-Einstein distribution with temperature $T=\Lambda_{\rm s}/\alpha_{\rm s}$. The scales $\Lambda_{\rm s}$ and $\Lambda$ are 
obtained from the integrals 
\begin{equation}
\frac{\Lambda\Lambda_{\rm s}}{\alpha_{\rm s}}\equiv -\int_0^\infty dp\, p^2
\frac{df}{dp}\,,\qquad \frac{\Lambda\Lambda_{\rm s}^2}{\alpha_{\rm s}^2}\equiv
\int_0^\infty dp\, p^2 f(1+f).  
\end{equation} 

Remarkably, in the regime where $f\gg 1$ ($f\sim 1/\alpha_{\rm s}$), all dependence on $\alpha_{\rm s}$ drops from the collision integral.
Taking moments of the collision integral with arbitrary powers of $p$, the typical collision time is given by $t_{\rm scat} = {\Lambda \over \Lambda_{\rm s}^2}$. 
The scattering time is itself a function of time and simple analysis of moments of the kinetic equation suggests that $ t_{\rm scat}\sim t$. For a fixed box, the energy conservation condition\footnote{The gluon number is not conserved because of the B-E condensate and/or inelastic number changing processes. We are also assuming here that the energy density of the condensate is small, as argued in ref.~[\refcite{Blaizot:2011xf}].} gives $ \Lambda_{s} \Lambda^3 \sim {\rm constant}$. 
With these two conditions, one obtains the temporal evolution of the two scales to be
\begin{equation}
  \Lambda_{\rm s} \sim Q_{s} \left( {t_0 \over t} \right)^{\frac{3}{7}}\;;\; 
 \Lambda \sim Q_{s} \left( {t \over t_0} \right)^{\frac{1}{7}} \, .
\end{equation}
The number density of gluons $n_{\rm g} \sim \Lambda^2 \Lambda_{s}$
decreases as $\sim (t_0/t)^{1/7}$.  The Debye mass $m = \sqrt{\Lambda_s \Lambda}$ decreases slowly in time as $\sim Q_{\rm s} (t_0/t)^{1/7}$. 
The thermalization time, determined from $\Lambda_{\rm s} \sim \alpha_{\rm s} \Lambda $, is 
$t_{\rm th} \sim {1 \over Q_{\rm s}}\left( {1\over \alpha_{\rm s}} \right)^{\frac{7}{4}}$.
Because the scale $\Lambda$ increases with time, so does the entropy density $s\sim \Lambda^3\sim Q_{\rm
  s}^3 (t/t_0)^{3/7}$. When $t=t_{\rm th}$, $s\sim Q_{\rm s}^3/\alpha_{\rm
  s}^{3/4}$, which is the equilibrium entropy $\sim T^3$.  Finally, becase quarks have a
phase space density of order 1 up to the scale $\Lambda$, the number density of quarks is $n_{\rm quarks} \sim \Lambda^3$. At $t\sim 1/Q_S$, this is 
suppressed compared to the number of gluons  $n_{\rm g}\sim \Lambda_{\rm s}\Lambda^2/\alpha_{\rm s}$ by $1/\alpha_{\rm s}$ . They become of the 
same order  when $\Lambda_{\rm s} \sim \alpha_{\rm s} \Lambda$, the thermalization time, and cannot be ignored at this time. 

We now consider the effect of longitudinal expansion, which can be included by adding a drift term to the left hand side of the kinetic equation. 
By integrating over momentum the kinetic equation multiplied by the energy one obtains
\begin{eqnarray} \label{eq:energy_expansion}
\partial_t \epsilon + \frac{\epsilon+P_{_L}}{t} =0\,,
\end{eqnarray}
where $\epsilon $ is the energy density, $\epsilon= \int {{d^3 p}\over {(2\pi)^3}}\,\omega_{\bf p} f_{\bf p}$, and $P_{_L}$ the longitudinal pressure,
$P_{_L}= \int {{d^3 p} \over {(2\pi)^3}} \frac{p_z^2}{\omega_{\bf p}} f_{\bf p}$. The effect of longitudinal expansion is observed by parameterizing
the longitudinal pressure in terms of energy density as $P_{_L} = \delta\, \epsilon$ ,where the multiplicative factor
$\delta$ can be in the range $[0,1/3]$; $\delta=0$ corresponds to the free streaming limit and $\delta=1/3$ to that of 
to ideal hydrodynamic expansion after isotropization. The assumption that $\delta$ is independent of time is a strong one which we make to understand how 
collisions redistribute momenta and thereby generate the shape of a thermal distribution for an expanding system. 

Within this framework,  $\epsilon_g(t) \sim \epsilon(t_0) \left(t_0\over t \right)^{1+\delta}$. Our previous estimate of the collision time is unchanged and can be combined 
with this expression to yield,
\begin{equation} \Lambda_{\rm s}\sim Q_{\rm s} \left(
  \frac{t_0}{t}\right)^{(4+\delta)/7},\qquad \Lambda\sim Q_{\rm s} \left(
  \frac{t_0}{t}\right)^{(1+2\delta)/7}.  
\end{equation} 
From these, one easily obtains the estimates of the gluon density, and of the Debye mass,
\begin{equation} \label{eq:gluons} n_{\rm g}\sim \frac{Q_{\rm
      s}^3}{\alpha_{\rm s}} \left(
    \frac{t_0}{t}\right)^{(6+5\delta)/7},\qquad m^2\sim Q_{\rm
    s}^2 \left(
    \frac{t_0}{t}\right)^{(5+3\delta)/7} \, .
\end{equation} 
The thermalization time, obtained as before from the condition
$\Lambda_{\rm s}=\alpha_{\rm s} \Lambda$, is given by $\left( \frac{t_{\rm th}}{t_0} \right)\sim
  \left(\frac{1}{\alpha_{\rm s}}\right)^{ \frac{7}{3-\delta}}$. Comparing this result to the fixed box result, we see that the expansion has the effect of delaying 
thermalization. (Formally, one recovers the fixed box result by setting $\delta=-1$, which corresponds to constant energy density).

For moderate values of the anisotropy (more precisely for $\delta>1/5$), a condensate can form, with number and energy density given by
\begin{equation} \label{eq:condensate}
n_{\rm c}\sim \frac{Q_{\rm s}^3}{\alpha_{\rm s}}\left( \frac{t_0}{t} \right)\left[ 1-\left( \frac{t_0}{t} \right)^{(-1+5\delta)/7}    \right]\,;\,\frac{\epsilon_{\rm c}}{\epsilon_{\rm g}}\sim  \left(  \frac{t_0}{t}\right)^{(5-11\delta)/14} ,
\end{equation}
The energy density of the condensate is subleading and decreases with increasing $t$. We should also note, as shown in ref.~[\refcite{Blaizot:2011xf}], that in a relaxation time 
approximation of the collision integral, solutions exist that correspond to a fixed time independent anisotropy. We shall now comment on the role of inelastic $2\leftrightarrow 3$ processes~[\refcite{Mueller:2006up}]. One can show that the inelastic scattering rate is parametrically the same order $t_{\rm inelastic} \sim {\Lambda \over \Lambda_{\rm s}^2} \sim t_{\rm scat}$. Because the $2\leftrightarrow 2$ transport equation for the overoccupied distribution ($\propto 1/\omega_{\bf p}$) provides a powerful source term for the Bose-Einstein condensate, it can exist as a transient state perhaps until close to thermalization when presumably inelastic scatterings become dominant. Finally, we should note that the result of ref.~[\refcite{Blaizot:2011xf}] for a fixed box coincides with that of ref.~[\refcite{Kurkela:2011ti}] for (among many scenarios considered) what we consider to be the physical scenario corresponding to the Glasma. For an expanding Glasma, the emphasis in ref.~[\refcite{Kurkela:2011ub}] is on plasma instabilities, while in our case strong elastic scattering is the dominant mechanism. Our results for thermalization do not require Bose-Einstein condensation; the possible creation of a transient condensate is plausible and may have interesting phenomenological consequences.  We anticipate that future non-perturbative non-equilibrium simulations outlined in ref.~[\refcite{Dusling:2011rz}] (and possible refinements thereoff [\refcite{Hatta:2011ky}]) will help clarify this issue.

\section*{Acknowledgments}

This work was supported under DOE Contract No. DE-AC02-98CH10886. I would like to thank Ian Balitsky, Alexei Prokudin and Anatoly Radyushkin for 
their organization of this interesting workshop.

\end{document}